\documentclass[prd,a4paper,nofootinbib,preprint]{revtex4-1} %
\usepackage{graphicx}
\usepackage{amsfonts}
\usepackage{amssymb}
\usepackage{slashed} 
\usepackage{multirow}
\usepackage[section]{placeins}
\usepackage{amsmath}
\usepackage{mathrsfs}
\usepackage{amsmath}
\usepackage{subfigure}
\usepackage{color}
\usepackage{hyperref}
\definecolor{oucrimsonred}{rgb}{0.6, 0.0, 0.0}
\definecolor{persianblue}{rgb}{0.11, 0.22, 0.73}
\definecolor{forestgreen}{rgb}{0.13,0.35,0.13}
 \hypersetup{colorlinks, citecolor=oucrimsonred, linkcolor=persianblue, urlcolor=oucrimsonred}

\begin{document}
%\preprint{CERN-TH-2016-091}
%%%%%%%%%%%%%%%%%%%%%%%%%%%%%%%%%%%%%%%%%%%%%%%%%%%%%%%%%%% FRONT PAGE
\title{The Scale Invariant Scotogenic Model: Dark Matter and the Scalar Sector}

\author{Rachik Soualah$^{1,2}$}
\email{rsoualah@sharjah.ac.ae, rsoualah@cern.ch}

\author{Amine Ahriche$^{1,2}$}
\email{ahriche@sharjah.ac.ae, amine.ahriche@cern.ch}

\affiliation{$^{1}$Department of Applied Physics and Astronomy, University of Sharjah, UAE.}
\affiliation{$^{2}$The International Center for Theoretical Physics (ICTP)- Strada Costiera, 11, I - 34151 Trieste Italy.}

\begin{abstract}
In this paper, we investigate the mutual impact between the dark matter
(DM) requirements and the scalar sector in the scale invariant (SI)
scotogenic model. The model is motivated by the neutrino mass and
DM within a classically SI framework. It is a SI generalization of
the scotogenic model, where the standard model (SM) is extended by
a real singlet, an inert scalar doublet and three Majorana singlet fermions, where the lightest one ($N_{1}$) could play the DM candidate role. In addition to the annihilation channels $N_{1}N_{1}\rightarrow\ell_{\alpha}\ell_{\beta},\nu_{\alpha}\bar{\nu}_{\beta}$,
the DM can be annihilated via few s-channel processes into SM particles, 
that are mediated by the Higgs/dilaton. This allows the new Yukawa
interactions, that are responsible for neutrino mass generation, to
take small values and therefore avoid the mass degeneracy between
the CP-even and CP-odd inert scalars unlike the case of the minimal
scotogenic model. In contrast to many Majorana DM models, the DM in
the SI-scotogenic model couples to the quarks at tree-level, and hence
the constraint from the direct detection experiments is very important
on the space parameter. The aim of this work is to investigate
the correlation between the DM requirements and the scalar sector in this model.
\end{abstract}
\maketitle

\section{Introduction}

The discovery of the Higgs boson on July 2012 provided a well established
explanation of the mass origin of the fermions and gauge bosons of
the standard model (SM). However, despite this great breakthrough,
many questions are still open, such as the origin of neutrino mass
and its smallness, the dark matter (DM) nature, that appears necessary
at the galactic scale. In addition to these unanswered questions,
it is so puzzling that most of the mass-dimension parameters in the
SM are of the order of $\mathcal{O}(100)$~$\mathrm{GeV}$. Therefore, we still
have too much to learn about the mechanism(s) of masses at the Universe.
In regards to this open issue, recent works have studied several extensions
of the SM that possess a scale-invariance (SI) symmetry, where the
electroweak symmetry breaking (EWSB) occurs as a quantum effect via
the radiative symmetry breaking a la Coleman-Weinberg~\cite{Coleman:1973jx}.
These models are phenomenologically rich, and generally predict new
$\mathrm{TeV}$ mass particles, that could be within the reach of the LHC.

A popular explanation of the smallness of the neutrino mass is provided by
the so-called seesaw mechanism~\cite{Gell}, where a hierarchy between
the charged leptons and neutrinos masses emerges due to the hierarchy
between the electroweak (EW) scale and the heavy singlet Majorana
fermion mass, that are added to the SM. The new high scale (Majorana
fermion mass) makes this scenario impossible to be probed at the current
and near future high energy colliders. Another attractive approach
is that the radiative neutrino mass is massless at tree level and
acquire a naturally small Majorana mass term at the loop level: one-loop~\cite{Zee:1985rj,Ma:1998dn},
two loops~\cite{Zee:1985id,Babu:1988ki,Aoki:2010ib,Guo:2012ne,Kajiyama:2013zla},
three loops~\cite{Krauss:2002px,Aoki:2008av,Gustafsson:2012vj,Ahriche:2014cda,Hatanaka:2014tba,Nishiwaki:2015iqa,Ahriche:2015wha,Ahriche:2015loa,Okada:2015hia,Nomura:2016ezz,Gu:2016xno,Cheung:2017efc,Dutta:2018qei},
and four loops~\cite{Nomura:2016fzs}. Some of these models address
the DM problem where a massive Majorana DM with a mass range from
$\mathrm{GeV}$ to $\mathrm{TeV}$ can play an important role as a
DM candidate~\cite{Ma:1998dn,Krauss:2002px,Aoki:2008av,Okada:2015hia},
in addition to the neutrino oscillation data. Furthermore, these models
predict interesting signatures at current/future collider experiments~\cite{Aoki:2016wyl,Ahriche:2014xra,Toma:2013zsa}
(for a review, see~\cite{Cai:2017jrq}).

Extending some of the $\nu$-DM motivated models by incorporating
the SI symmetry, makes the model phenomenology modified and richer~\cite{Foot:2007ay}.
Indeed, one can emphasize that by considering the SI symmetry the
issues of ESWB, neutrino mass and DM are solved all together at the
EW scale. For example, the models in~\cite{Ahriche:2016cio} and
\cite{Ahriche:2015loa} are SI generalizations of a one-loop (scotogenic~\cite{Ma:2006km})
and a three-loop (KNT~\cite{Krauss:2002px}) model, where the lightest
Majorana fermion plays the DM role. In addition to the DM annihilation
channel $N_{1}N_{1}\rightarrow\ell_{\alpha}\ell_{\beta},\,\nu_{\alpha}\bar{\nu}_{\beta}$
in the scotogenic model (and similarly with the SI-KNT model~\cite{Ahriche:2015loa}),
other channels such as $N_{1}N_{1}\rightarrow h_{i}h_{k},\,VV,q\bar{q}$,
are possible to take place due to the DM coupling with the Higgs/dilaton
in the SI extensions. Here, $V=W,Z$ are the SM gauge bosons, and
$h_{i}$ could be the Higgs $H$ and/or the the lighter CP-even eigenstate
$D$, that is called the dilaton in the SI models. It is strictly massless
at tree-level and acquires its mass via the loop corrections. In the
non SI models (for example~\cite{Krauss:2002px} and~\cite{Ma:2006km}),
for a given DM mass $M_{1}$, the relic density dictates the values
of the new Yukawa couplings ($g_{i\alpha}$) that couple the left-handed
lepton doublets (or the charged right-handed leptons in the case of
KNT model) and the inert doublet (charged singlet) scalar. This makes
the values of these couplings ($g_{i\alpha}$) too constrained since
they should be also in agreement with the neutrino oscillation data~\cite{Fukuda:2001nj},
and the lepton flavor violating (LFV) constraints~\cite{PDG2021}.
In the SI framework, the non negligible contributions of the new DM
annihilation channels $N_{1}N_{1}\rightarrow SS,\,VV,q\bar{q}$ could
relax the bounds on these new couplings and make the parameters space
(especially the range of the couplings $g_{i\alpha}$) much larger.
The new DM-Higgs/dilaton interactions that allow the new DM annihilation
channels will lead to a tree-level DM-nucleon scattering cross section,
contrary to the non SI models where it is a pure one-loop effect.
Consequently, this makes the effect of the requirement of both the
direct detection (DD) constraints and the relic density, very important
on the scalar sector within the SI models. For this reason, investigating
these effects is the motivation of our work by considering the model~\cite{Ahriche:2016cio}
as an example. In the non-SI scotogenic model with fermionic DM~\cite{Ahriche:2017iar},
the neutrino mass smallness is achieved by imposing a mass degeneracy
between the neutral inert CP-even and CP-odd scalars, since relatively
large values for the $g_{i\alpha}$ couplings are dictated by the
DM relic density. This implies that one of the scalar quartic couplings
must be suppressed $\lambda_{5}=\mathcal{O}(10^{-10})$. Nonetheless,
this fine-tuning could be avoided by extending the model with a
real singlet that is charged under a global $Z_{4}/Z_{2}$ symmetry~\cite{Ahriche:2020pwq}.
Here, we will investigate the possibility of non-suppressed $\lambda_{5}$
value within the SI framework due to the existence of new DM annihilation
channels ($N_{1}N_{1}\rightarrow SS,\,VV,q\bar{q}$).

The paper is organized as follows. Section~\ref{sec:The-Model} is
devoted to describe the SI scotogenic $\nu$-DM model, in addition
to the EWSB description within the SI framework. In section~\ref{sec:DM},
the DM relic density is estimated via an explicit and exact calculation
of the cross section of all channels. In addition, to impose the DD
bounds, the DM nucleon cross section formula is given. In section~\ref{sec:NR},
we perform a random scan for the model parameters space to probe all
the possible correlation aspects between the scalar sector and the
DM requirements. Our conclusions are given in section~\ref{sec:Conc}.

\section{The Scale Invariant Scotogenic Model\label{sec:The-Model}}

Here, the SM is extended by one inert doublet scalar, $S\sim(1,2,1)$,
three singlet Majorana fermions $N_{i}\sim(1,1,0)$, and one real
neutral singlet scalar $\phi\sim(1,1,0)$ to assist the radiative
EWSB. The model is assigned by a global $Z_{2}$ symmetry $\{S,\,N_{i}\}\rightarrow\{-S,\,-N_{i}\}$,
where all other fields being $Z_{2}$-even, in order to make the lightest
$Z_{2}$-odd field $N_{1}\equiv N_{\text{DM}}$ stable, hence it 
plays the DM candidate role. The Lagrangian contains the following
terms 
\begin{equation}
\mathcal{L}\supset-\;\{g_{i\alpha}\overline{N_{i}^{c}}S^{\dagger}L_{\alpha}+\mathrm{h.c}\}-\frac{1}{2}y_{i}\phi\overline{N_{i}^{c}}\,N_{i}-V(H,S,\phi),\label{L:Ma}
\end{equation}
where the inert doublet can be presented as $S^{T}=\Big(S^{+}\,,\,(S^{0}+i\,A^{0})/\sqrt{2}\Big)\sim(1,2,1)$,
$L_{\beta}$ and $\ell_{\alpha R}$ the left-handed lepton doublet
and right-handed leptons; the Greek letters label the SM flavors,
$\alpha,\,\beta\in\{e,\,\mu,\,\tau\}$, $g_{i\alpha}$ and $y_{i}$
are new Yukawa couplings. The most general SI scalar potential that
obeys the $Z_{2}$ symmetry is given by 
\begin{align}
V_{0}(H,\,S,\,\phi) & =\frac{1}{6}\lambda_{H}(\left|H\right|^{2})^{2}+\frac{\lambda_{\phi}}{24}\phi^{2}+\frac{\lambda_{S}}{2}|S|^{4}+\frac{\omega_{1}}{2}|H|^{2}\phi^{2}+\frac{\omega_{2}}{2}\,\phi^{2}|S|^{2}+\lambda_{3}\,|H|^{2}|S|^{2}\nonumber \\
 & +\lambda_{4}\,|H^{\dagger}S|^{2}+\{\frac{\lambda_{5}}{2}(H^{\dagger}S)^{2}+h.c.\}\label{V:Ma}
\end{align}

\begin{figure}[t]
\begin{centering}
\includegraphics[width=0.4\textwidth]{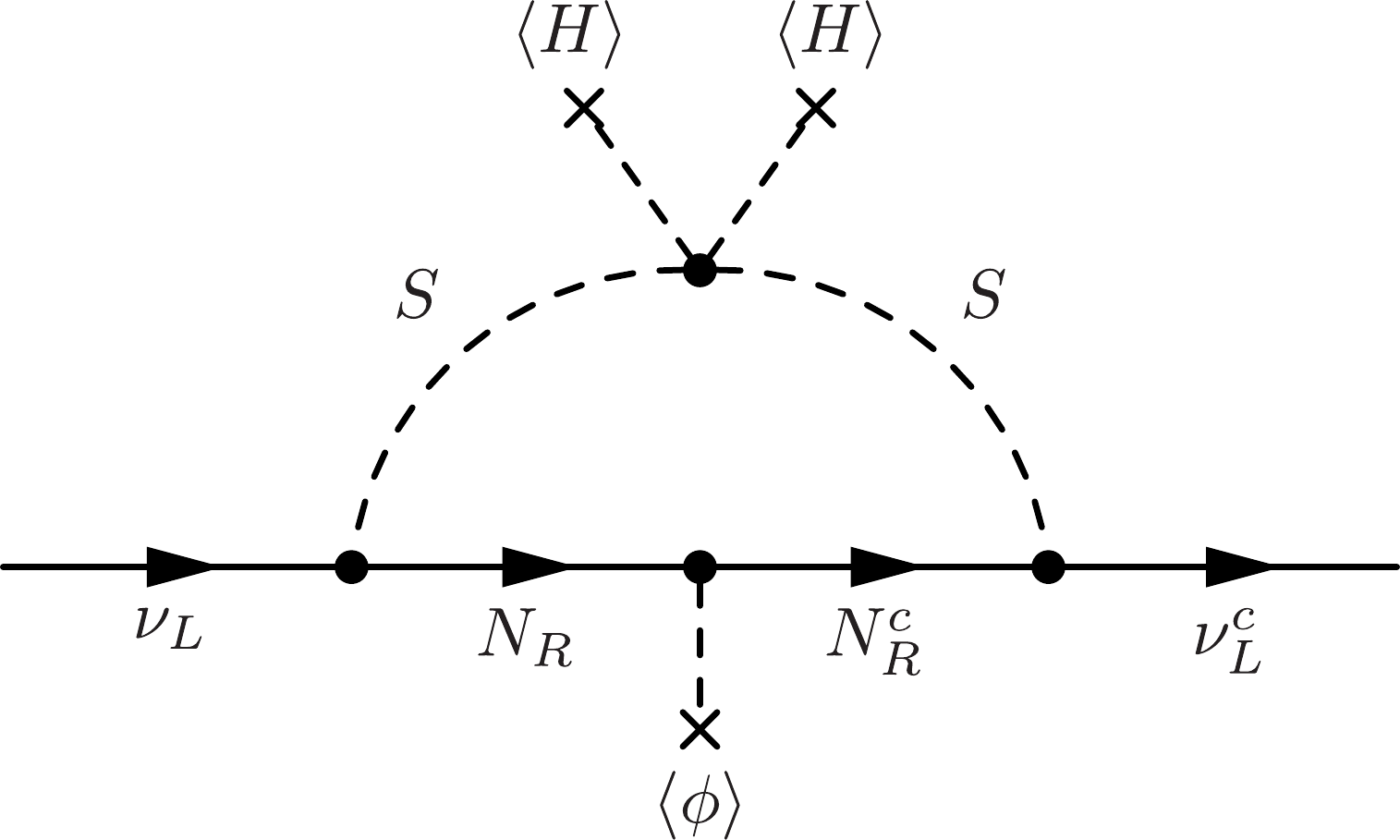} 
\par\end{centering}
\caption{The neutrino mass is generated in the SI-scotogenic model at one-loop
level.}
\label{fig:SI3l} 
\end{figure}

The first term in (\ref{L:Ma}) and the last term in (\ref{V:Ma})
are responsible for generating neutrino mass via the one-loop diagrams
as illustrated in Fig.~\ref{fig:SI3l}.

The neutrino mass matrix elements are given by~\cite{Ma:2006km}
\begin{equation}
m_{\alpha\beta}^{(\nu)}=\sum_{i}\frac{g_{i\alpha}g_{i\beta}M_{i}}{16\pi^{2}}\left[\frac{m_{S^{0}}^{2}}{m_{S^{0}}^{2}-M_{i}^{2}}\log\frac{m_{S^{0}}^{2}}{M_{i}^{2}}-\frac{m_{A^{0}}^{2}}{m_{A^{0}}^{2}-M_{i}^{2}}\log\frac{m_{A^{0}}^{2}}{M_{i}^{2}}\right].\label{eq:MaNU}
\end{equation}
The neutrino mass matrix elements in (\ref{eq:MaNU}) can be written
as $m_{\alpha\beta}^{(\nu)}=\sum_{i}g_{i\alpha}g_{i\beta}\Lambda_{i}=\left(g^{T}.\varLambda.g\right)_{\alpha\beta}$,
which permits us to write the new Yukawa couplings according to the
Casas-Ibarra parameterization as~\cite{Casas:2001sr} 
\begin{equation}
g=D_{\sqrt{\varLambda^{-1}}}RD_{\sqrt{m_{\nu}}}U_{\nu}^{T},
\end{equation}
with $D_{\sqrt{\varLambda^{-1}}}=\textrm{diag\{\ensuremath{\varLambda_{1}^{-1/2}},\ensuremath{\varLambda_{2}^{-1/2}},\ensuremath{\varLambda_{3}^{-1/2}}\}},\,D_{\sqrt{m_{\nu}}}=\textrm{diag}\{m_{1}^{1/2},m_{2}^{1/2},m_{3}^{1/2}\}$,
$R$ is an arbitrary $3\times3$ orthogonal matrix, $m_{i}$ represents
the neutrino mass eigenstates and $U_{\nu}$ is the Pontecorvo-Maki-Nakawaga-Sakata
(PMNS) mixing matrix. These couplings are subject of the LFV bounds
on the branching ratios of $\ell_{\alpha}\rightarrow\ell_{\beta}\gamma$
and $\ell_{\alpha}\rightarrow\ell_{\beta}\ell_{\beta}\ell_{\beta}$~\cite{Toma:2013zsa}.

In this model, the DM candidate could be fermionic (the lightest Majorana
fermion, $N_{1}$) or a scalar (the lightest among $S^{0}$ and $A^{0}$).
In the case of a scalar DM, the situation matches the inert doublet
model case~\cite{Banerjee:2021oxc}, where the co-annihilation effect
should be considered in order to have viable parameters space. For
Majorana DM case, the Yukawa couplings $g_{i\alpha}$ values are constrained
by the relic density, and therefore the neutrino mass smallness should
be achieved by the $S^{0}-A^{0}$ mass degeneracy, i.e., imposing
a very small values for\textbf{ $\lambda_{5}=\mathcal{O}(10^{-10})$}.

After the EWSB, the CP-even neutral scalars acquire VEV's as 
\begin{equation}
H\rightarrow\frac{\upsilon+h}{\sqrt{2}}\begin{pmatrix}0\\
1
\end{pmatrix},\,\,\,\phi\rightarrow\frac{x+\phi}{\sqrt{2}},\label{vev}
\end{equation}
and hence give masses to all model fields, where we get two CP-even
eigenstates 
\begin{equation}
\begin{pmatrix}H\\
D
\end{pmatrix}=\begin{pmatrix}c_{\alpha}~-s_{\alpha}\\
s_{\alpha}~c_{\alpha}
\end{pmatrix}\begin{pmatrix}h\\
\phi
\end{pmatrix},\label{scMix}
\end{equation}
where $H$ denotes the 125 $\mathrm{GeV}$ Higgs, $D$ is the dilaton scalar and
$\alpha$ is the Higgs-dilaton mixing angle, which is defined at tree-level
as $s_{\alpha}=\sin\alpha=\upsilon/\sqrt{\upsilon^{2}+x^{2}}$ and
$c_{\alpha}=\cos\alpha=x/\sqrt{\upsilon^{2}+x^{2}}$. In the SI approach,
the EWSB is triggered by the the radiative corrections. Then, the
one-loop effective potential in (\ref{V:Ma}) can be expressed in
the $\overline{DR}$ scheme as 
\begin{eqnarray}
V^{1-\ell}(h,\phi) & = & \frac{\lambda_{H}+\delta\lambda_{H}}{24}h^{4}+\frac{\lambda_{\phi}+\delta\lambda_{\phi}}{24}\phi^{4}+\frac{\omega+\delta\omega}{4}h^{2}\,\phi^{2}\nonumber\\
 & & + \frac{1}{64\pi^{2}}\sum_{i}n_{i}m_{i}^{4}(h,\phi)\left[\log\frac{m_{i}^{2}(h,\phi)}{\Lambda^{2}}-\frac{3}{2}\right],\label{eq:V}
\end{eqnarray}
where $\delta\lambda_{H},\,\delta\lambda_{\phi},\,\delta\omega$ are
counter-terms, $n_{i}$ and $m_{i}^{2}(h,\phi)$ are the field multiplicity
and field dependant masses, respectively. Here $\Lambda=m_{h}=125.18\,\mathrm{GeV}$
is the renormalization scale and $m_{h}$ is the measured Higgs mass. Here,
we choose the counter-terms $\delta\lambda_{H},\,\delta\lambda_{\phi}$
and $\delta\omega$, where the minimum $\{h=\upsilon,\phi=x\}$ is
still the vacuum at one-loop; and the Higgs mass at one-loop must
correspond to the measured value. The field dependent masses for the
gauge bosons and fermions are the same as in the SM. The field dependent
masses of the Goldstone and the Majorana singlets are $m_{\chi}^{2}=\frac{1}{6}\lambda_{H}h^{2}+\frac{1}{2}\omega\phi^{2}$
and $M_{i}^{2}=y_{i}^{2}\phi^{2}$, respectively. The Higgs-dilation
eigenmasses are the eigenvalues of the matrix whose elements are $m_{hh}^{2}=\frac{1}{2}\lambda_{H}h^{2}+\frac{1}{2}\omega\phi^{2}$,
$m_{\phi\phi}^{2}=\frac{1}{2}\lambda_{\phi}\phi^{2}+\frac{1}{2}\omega h^{2}$
and $m_{h\phi}^{2}=\omega h\phi$, and the inert field dependant masses
are given by $m_{S^{\pm}}^{2}=\frac{1}{2}\omega_{2}\phi^{2}+\frac{1}{2}\lambda_{3}h^{2},~m_{S^{0},A^{0}}^{2}=m_{S^{\pm}}^{2}+\frac{1}{2}(\lambda_{4}\pm\lambda_{5})h^{2}$.

Since the dilaton acquires mass via radiative effects, then its mass
squared should be positive $m_{D}^{2}>0$, which is basically guaranteed
by the vacuum stability conditions. The vacuum stability could be
ensured by imposing the coefficients of the term $\varphi^{4}\log\varphi$
to be positive rather than the coefficients of the term $\varphi^{4}$, where $\varphi$ refers
to any direction in the $h-\phi$ plan. These conditions are translated
into 
\begin{equation}
\begin{array}{c}
-12y_{t}^{4}+\tfrac{9}{4}g_{2}^{4}+\tfrac{3}{4}g_{1}^{4}+\frac{4}{3}\lambda_{H}^{2}+\frac{4}{3}\omega_{1}^{2}+2\lambda_{3}^{2}+(\lambda_{3}+\lambda_{4}+\lambda_{5})^{2}+(\lambda_{3}+\lambda_{4}-\lambda_{5})^{2}>0,\\
\lambda_{\phi}^{2}+4\omega_{1}^{2}+4\omega_{2}^{2}-8(y_{1}^{4}+y_{2}^{4}+y_{3}^{4})>0.
\end{array}\label{eq:Ma-VSt}
\end{equation}

As mentioned earlier, the EWSB is triggered by the radiative corrections,
where the Higgs/dilaton sector has a similar structure for all the
SI $\nu$-DM models. However, there may be some few differences due
to the nature and multiplicity of the new fields. Indeed, the new
model couplings are not fully free and are severely constrained by
many conditions such as the measured Higgs mass and the vacuum stability.
Since the dilaton squared mass is purely radiative, we try to quantify
the radiative effects by the values of both of the dilation squared
mass and the one-loop correction to Higgs/dilaton mixing 
\begin{equation}
\Delta_{s_{\alpha}}=\frac{(s_{\alpha})_{1-loop}-(s_{\alpha})_{tree-level}}{(s_{\alpha})_{tree-level}}.\label{eq:DS}
\end{equation}
These two quantities, in addition to the singlet VEV controls the
new contribution of the DM annihilation via the interactions with the
Higgs/dilaton, i.e., via the channels $N_{1}N_{1}\rightarrow h_{i}h_{k},\,VV,q\bar{q}$
with $h_{i}=H,D$.

One has to mention that for other regions of space parameters, the light CP-even could match the SM-like Higgs, where the Higgs mass 
is generated fully radiative~\cite{Ahriche:2021frb}. In this setup, the radiative effects push simultaneously the light CP-even mass to match the measured Higgs mass; and the scalar mixing to be in agreement with the Higgs signal strength measurements at the LHC~\cite{ATLAS:2016neq}. 
In addition, this scenario is in agreement with all the previously mentioned constraints~\cite{Ahriche:2021frb}.

\section{Dark Matter\label{sec:DM}}

In this section, we discuss the relic density estimation approach
and give the exact cross section formulas for the different annihilation
channels. Then, we give the DM-nucleon scattering cross section that
is subject of DD experiments. 

\textbf{DM Relic Density}: the DM candidate is the lightest Majorana
fermion ($N_{1}$), where there are many possible annihilation channels,
that can be classify into two categories according to the DM interactions:
(1) the annihilation via the new Yukawa interactions ($g_{i\alpha}$),
i.e., into $N_{1}N_{1}\rightarrow\ell_{\alpha}\ell_{\beta},\,\nu_{\alpha}\bar{\nu}_{\beta}$;
and (2) the annihilation via the interactions with the Higgs/dilaton,
i.e., $N_{1}N_{1}\rightarrow h_{i}h_{k},\,VV,q\bar{q}$, where $S_{i}=H,\,D$
and $V=W,Z$. Clearly, for suppressed Higgs-dilation mixing $s_{\alpha}$,
the situation matches the non SI models, i.e., the DM annihilation
must fully achieved via the new Yukawa interactions. In the opposite
case, the DM annihilation could be dominated by all channels (or simply
by just one channel) that are mediated by the Higgs/dilaton. This
depends on the DM mass $M_{1}$, dilaton mass and the Higgs-dilation
mixing $s_{\alpha}$.

\begin{figure}[t]
\begin{centering}
\includegraphics[width=1.0\textwidth]{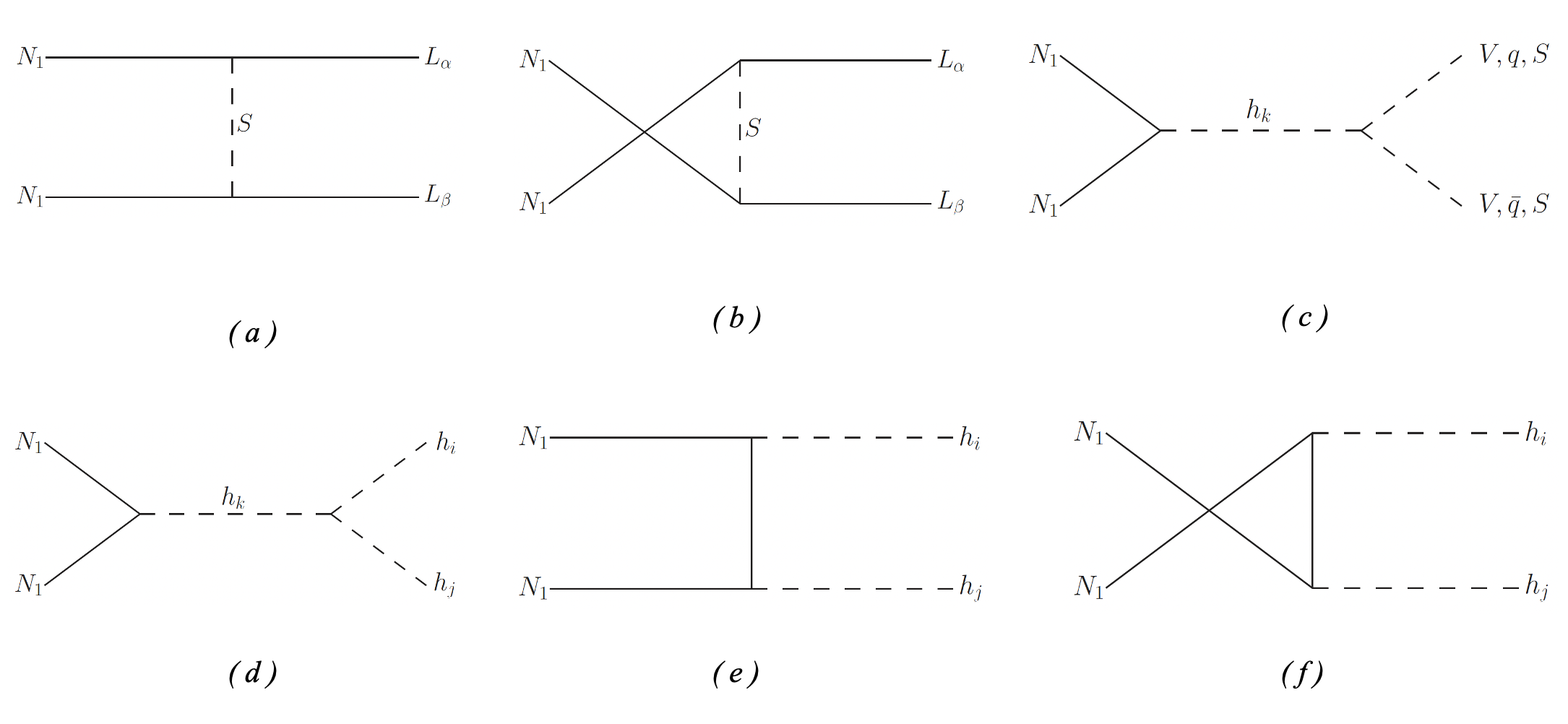} 
\par\end{centering}
\caption{Different DM annihilation channels in the SI-scotogenic model. The cross section in (\ref{eq:XS1}) is estimated based on the diagrams (a) and (b), while diagram (c) yields the cross section of the SM channels in (\ref{csSM}). The Higgs/dilaton annihilation cross section in (\ref{HHDD}) is computed using the diagrams (d), (e) and (f).}
\label{DM-ahn} 
\end{figure}

In what follows, we show how could the DM relic density be estimated
at the freeze-out temperature~\cite{Srednicki:1988ce}. When the
temperature of the Universe drops below the DM mass, the DM decouples
from the thermal bath and then preserves the DM particles number.
In other words, after the freeze-out the ratio $n_{DM}/n_{s}$ remains
constant during the Universe expansion, where $n_{DM}$ and $n_{s}$
are the DM number and entropy densities, respectively. The estimated
relic density must match the Planck observation~\cite{Aghanim:2018eyx}
\begin{equation}
\Omega_{{\rm DM}}h^{2}=0.120\pm0.001\,,\label{eq:omegah}
\end{equation}
where $h$ is the reduced Hubble constant and $\Omega_{{\rm DM}}$
denotes the DM energy density scaled by the critical density. Up to
a very good approximation, the cold DM relic
abundance of the WIMP scenario is given by~\cite{Srednicki:1988ce} 
\begin{equation}
\Omega_{N_{1}}h^{2}\simeq\frac{(1.07\times10^{9})x_{F}}{\sqrt{g_{\ast}}M_{pl}(\mathrm{GeV})\left\langle \sigma(N_{\text{1}}\ N_{\text{1}})\upsilon_{r}\right\rangle },
\end{equation}
where $x_{F}=M_{1}/T_{F}$ represents the freeze-out temperature,
that can be determined iteratively from the equation 
\begin{equation}
x_{F}=\log\left(\sqrt{\frac{45}{8}}\frac{M_{1}M_{pl}\left\langle \sigma(N_{1}N_{1})\upsilon_{r}\right\rangle }{\pi^{3}\sqrt{g_{\ast}x_{F}}}\right).
\end{equation}
Here, $\upsilon_{r}$ denotes the relative velocity, $M_{pl}$ is
the Plank mass, $g_{\ast}$ counts the effective degrees of freedom
of the relativistic fields in equilibrium, and 
\begin{align}
\left\langle \sigma(N_{1}N_{1})\upsilon_{r}\right\rangle & =\sum_{X}\left\langle \sigma(N_{1}N_{1}\rightarrow X)\upsilon_{r}\right\rangle =\sum_{X}\int_{4M_{1}^{2}}^{\infty}ds~\sigma_{N_{1}N_{1}\rightarrow X}(s)\frac{\left(s-4M_{1}^{2}\right)}{8TM_{1}^{4}K_{2}^{2}\left(\frac{M_{1}}{T}\right)}\sqrt{s}K_{1}\left(\frac{\sqrt{s}}{T}\right),
\end{align}
represents the total thermally averaged annihilation cross section.
We have $s$ is the Mandelstam variable, $K_{1,2}$ are the modified
Bessel functions and $\sigma_{N_{1}N_{1}\rightarrow X}(s)$ is the
partial annihilation cross due to the channel $N_{1}N_{1}\rightarrow X$,
at the CM energy $\sqrt{s}$, where the possible channels are shown
in Fig.~\ref{DM-ahn}.

We provide here the exact formulas for the partial annihilation cross
sections. For the channels $N_{1}N_{1}\rightarrow\ell_{\alpha}\ell_{\beta},\,\nu_{\alpha}\bar{\nu}_{\beta}$,
we have~\cite{Lalili,Ahriche:2020pwq} 
\begin{align}
\sigma\upsilon_{r} & =\frac{1}{32\pi s^{2}}\sum_{\alpha,\beta}\sum_{X,Y}\left|\eta_{X}\eta_{Y}\,g_{1\alpha}g_{1\beta}^{*}\right|^{2}\,\lambda(s,m_{\alpha}^{2},m_{\beta}^{2})\left\{ \mathcal{R}(Q_{X},Q_{Y},T_{+},T_{-},B)+\right.\nonumber \\
 & \left.-\frac{M_{1}^{2}(s-m_{\alpha}^{2}-m_{\beta}^{2})}{B(Q_{X}+Q_{Y})}\log\left|\frac{(Q_{X}+B)(Q_{Y}+B)}{(Q_{X}-B)(Q_{Y}-B)}\right|\right\},\label{eq:XS1}\\
Q_{X} & =\frac{1}{2}(s+2m_{X}^{2}-2M_{1}^{2}-m_{\alpha}^{2}-m_{\beta}^{2}),\,T_{\pm}=\frac{1}{2}(s\pm m_{\alpha}^{2}\mp m_{\beta}^{2}),\nonumber \\
B & =\frac{1}{2s}\lambda(s,m_{\alpha}^{2},m_{\beta}^{2})\lambda(s,M_{1}^{2},M_{1}^{2}),\,\lambda(x,y,z)=\sqrt{(x-y-z)^{2}-4yz},\nonumber\\
\mathcal{R}(r,t,w,q,\eta) & =2-\frac{(t-w)(t-q)}{\eta\,(t-r)}\log\left|\frac{t+\eta}{t-\eta}\right|-\frac{(r-w)(r-q)}{\eta\,(r-t)}\log\left|\frac{r+\eta}{r-\eta}\right|.\nonumber 
\end{align}
For the channel $N_{1}N_{1}\rightarrow\ell_{\alpha}\ell_{\beta}$,
we have: $m_{\alpha}=m_{\ell_{\alpha}},\,m_{\beta}=m_{\ell_{\beta}}$, 
$\eta_{S^{\pm}}=i$ and $\{X,Y\}=\{S^{\pm},S^{\pm}\}$. While, for
the channel $N_{1}N_{1}\rightarrow\nu_{\alpha}\bar{\nu}_{\beta}$,
we have: $m_{\alpha}=m_{\beta}=0,\,\{X,Y\}=\{S^{0},S^{0}\},\{S^{0},A^{0}\},\{A^{0},A^{0}\}$
and $\eta_{S^{0}}=\frac{i}{_{\sqrt{2}}},\,\eta_{A^{0}}=\frac{1}{_{\sqrt{2}}}$.

For the channels that are mediated by the Higgs/dilaton ($N_{1}N_{1}\rightarrow X\bar{X},\,X=W,Z,b,t$),
one can write the cross section as~\cite{Ahriche:2016cio} 
\begin{equation}
\sigma(N_{1}N_{1}\rightarrow X\bar{X})\upsilon_{r}=8\sqrt{s}s_{\alpha}^{2}c_{\alpha}^{2}y_{1}^{2}\left\vert \frac{1}{s-m_{H}^{2}+im_{H}\Gamma_{H}}-\frac{1}{s-m_{D}^{2}+im_{D}\Gamma_{D}}\right\vert ^{2}\Gamma_{H\rightarrow X\bar{X}}(m_{H}\rightarrow\sqrt{s}),\label{csSM}
\end{equation}
where $\Gamma_{H\rightarrow X\bar{X}}(m_{H}\rightarrow\sqrt{s})$\ is
the total SM Higgs decay width with the Higgs mass replaced by $m_{H}\rightarrow\sqrt{s}$.
Here, $\Gamma_{H,D}$ are the Higgs and dilaton total decay widths,
respectively. For simplicity, we take in our numerical
scan $\Gamma_{H}\simeq\Gamma_{h}^{SM}$ and $\Gamma_{D}\simeq0$.
For the annihilation into Higgs/dilaton, we have the exact cross section
formula at CM energy $\sqrt{s}$ given by 
\begin{align}
\sigma(N_{1}N_{1}\rightarrow h_{i}h_{j})\upsilon_{r} & =\frac{1}{8\pi s}\left\{ R+\frac{W}{2B}\log\left|\frac{A-B}{A+B}\right|+\frac{Z}{A^{2}-B^{2}}\right\} ,\label{HHDD}
\end{align}
with $h_{i}=H,D$ and
\begin{align*}
R & =\frac{1}{4}y_{1}^{2}(s-M_{1}^{2})|Q|^{2},\,W=\frac{1+\delta_{ij}}{2}q_{i}q_{j}y_{1}^{3}(s-M_{1}^{2})M_{1}\left(\Re(Q)+\frac{\delta_{ij}q_{i}q_{j}y_{1}M_{1}}{(m_{h_{i}}^{2}+m_{h_{j}}^{2}-s)}\right),\\
Z & =\frac{1+\delta_{ij}}{4}q_{i}^{2}q_{j}^{2}y_{1}^{4}M_{1}^{2}(s-M_{1}^{2}),\,Q=\frac{-s_{h}\lambda_{Hij}}{s-m_{H}^{2}+im_{H}\Gamma_{H}}+\frac{c_{h}\lambda_{ijD}}{s-m_{D}^{2}+im_{D}\Gamma_{D}},\\
A & =\frac{1}{2}(s-m_{h_{i}}^{2}-m_{h_{j}}^{2}),\,B=\frac{1}{2s}\lambda(s,M_{1}^{2},M_{1}^{2})\lambda(s,m_{h_{i}}^{2},m_{h_{j}}^{2}),
\end{align*}
where $\lambda_{ijk}$ are the Higgs/dilaton triple couplings and
\{$q_{H}=-s_{\alpha},\,q_{D}=c_{\alpha}$\}.

\vspace{1cm}

\textbf{DM Direct Detection}: in the SI $\nu-$DM models, the sensitivity 
to the direct-detection experiments could be due to the interactions
between the DM and quarks via the mediation of the Higgs/dilaton.
The effective low-energy Lagrangian of this interaction can be written
as 
\begin{equation}
\mathcal{L}_{N_{1}-q}^{(eff)}=-\frac{1}{2}s_{\alpha}c_{\alpha}y_{q}y_{1}\left[\frac{1}{m_{H}^{2}}-\frac{1}{m_{D}^{2}}\right]\,\bar{q}q\,\overline{N}_{\text{1}}^{c}N_{1},
\end{equation}
where $y_{q}$ is the light quark Yukawa coupling. Consequently,
the effective nucleon-DM interaction can be written as 
\begin{equation}
\mathcal{L}_{N_{1}-\mathcal{N}}^{(eff)}=\frac{s_{\alpha}c_{\alpha}(m_{\mathcal{N}}-\frac{7}{9}m_{\mathcal{B}})M_{1}}{x\,\upsilon}\left[\frac{1}{m_{h}^{2}}-\frac{1}{m_{D}^{2}}\right]\mathcal{\bar{N}N}\overline{N}_{\text{1}}^{c}N_{\text{1}},
\end{equation}
where $m_{\mathcal{N}}$ is the nucleon mass and $m_{\mathcal{B}}$
the baryon mass in the chiral limit~\cite{He:2008qm}. This leads
to the nucleon-DM spin-independent elastic cross section in the chiral
limit~\cite{Ahriche:2016cio} 
\begin{equation}
\sigma_{\det}=\frac{c_{\alpha}^{2}s_{\alpha}^{2}m_{\mathcal{N}}^{2}(m_{\mathcal{N}}-\frac{7}{9}m_{\mathcal{B}})^{2}M_{1}^{4}}{\pi\upsilon^{2}x^{2}(M_{1}+m_{\mathcal{B}})^{2}}\left[\frac{1}{m_{h}^{2}}-\frac{1}{m_{D}^{2}}\right]^{2}.\label{eq:DD}
\end{equation}
In what follows, we will consider the recent upper bound reported
by Xenon 1T experiment~\cite{XENON:2018voc}. In addition, we compare
our results with the projected sensitivities for the future proposed
experiments: PandaX-4t~\cite{PandaX:2018wtu}, LUX-Zeplin~\cite{LUX-ZEPLIN:2018poe},
XENONnT with 20 ton-yr exposure~\cite{XENON:2020kmp} and DARWIN~\cite{DARWIN:2016hyl}.

\section{Numerical Results and Discussion\label{sec:NR}}

This model is subject to many theoretical and experimental constraints.
Here, we will be interested in all phenomenological and experimental
aspects of the Higgs/dilaton interactions and their correlation with
DM relic density and DD, which may imply some constraints on the neutrino
mass and the LFV processes. Then, in our numerical scan, we consider
the following constraints: perturbativity, perturbative unitarity,
the different Higgs decay channels (di-photon, invisible and undetermined),
the electroweak precision tests, LEP negative searches for light scalar
(in our case, it applies to the cross section of $e^{-}e^{+}\rightarrow ZD$),
and the Higgs signal strength at the LHC $\mu_{{\rm tot}}\geq0.89$
at 95\%~CL~\cite{ATLAS:2016neq}. The latter constraint (Higgs signal strength at the LHC~\cite{ATLAS:2016neq})
can be expressed as $\mu_{{\rm tot}}=c_{\alpha}^{2}\times(1-\mathcal{B}_{BSM})\geq0.89$,
with $c_{\alpha}$ being the Higgs/dilaton mixing angle and $\mathcal{B}_{BSM}$
is the branching ratio of any non-SM Higgs decay channel (could be
the invisible channel like $H\rightarrow N_{1}N_{1}$ or/and an undermined
channel like $H\rightarrow DD$), which are constrained as $\mathcal{B}_{BSM}\leq0.47$~\cite{Aad:2019mbh}.
For a given value of $\mathcal{B}_{BSM}$, the condition $\mu_{{\rm tot}}\geq0.89$
can be translated at tree-level into a direct constraint on the singlet
VEV value as: $x>700.36\,\mathrm{GeV},\,600\,\mathrm{GeV},\,480\,\mathrm{GeV},\,335\,\mathrm{GeV},\,233\,\mathrm{GeV}$
for $\mathcal{B}_{BSM}=0,\,0.038,\,0.11,\,0.27,\,0.469$, respectively.
However, one has to mention that in the SI models $\mathcal{B}_{BSM}$
is likely to take values much smaller than the experimental bound
due to the facts that the Higgs invisible decay branching ratio is
$s_{\alpha}^{2}$ proportional, and the triple coupling $\lambda_{HDD}$
is suppressed since it has only a pure one-loop contribution. Here,
we consider the input parameters ranges 
\begin{equation}
\begin{array}{c}
113.5\,\mathrm{GeV}<m_{S^{\pm}}<1\,\mathrm{TeV},\,10\,\mathrm{GeV}<m_{S^{0},A^{0}}<1\,\mathrm{TeV},\,6\,\mathrm{GeV}<M_{1}<1\,\mathrm{TeV},\\
1\,\mathrm{GeV}<m_{D}<100\,\mathrm{GeV},\,y_{i}^{2},\,\left|\lambda_{i}\right|\,\left|g_{i\alpha}\right|^{2}<4\pi,
\end{array}\label{eq:free}
\end{equation}
where $\lambda_{i}$ denotes all the couplings in (\ref{V:Ma}). In
the absence of BSM decay channels for the Higgs, the constraints $\mu_{{\rm tot}}\geq0.89$
implies that the singlet VEV is larger than $x>700.36\,\mathrm{GeV}$,
i.e., $x^{2}/(\upsilon^{2}+x^{2})\geq0.89$. However, by considering
the radiative correction in the mixing estimation, the singlet VEV
could smaller as $x\geq\,\mathrm{GeV}$. This is understood from
the fact that the radiative corrections could be large so that the
dilaton mass, which is a pure radiative effect, could reach values
as $m_{D}\sim100\,\,\mathrm{GeV}$, and the ratio $\Delta_{s_{\alpha}}$
in (\ref{eq:DS}) lies between $-1500\%<\Delta_{s_{\alpha}}<1500\%$.
This means that the scalar mixing could be dominated by the radiative corrections, and hence in our analysis, we will consider the one-loop value of the mixing $s_{\alpha}$.

The existence of the Higgs/dilaton coupling to the Majorana DM candidate,
$N_{1}$ in the SI models, leads to an addition contribution to the
DM annihilation cross section via the channels $N_{1}N_{1}\rightarrow h_{i}h_{k},\,VV,q\bar{q}$,
with $h_{i}$ denotes the Higgs/dilaton and $V$ denotes the gauge
bosons. This new contribution could be either small or dominant depending
on the parameters $m_{D},~M_{1}$ and the mixing $\sin\alpha$. Therefore,
such a non-negligible new contribution makes the contribution of the
channels $N_{1}N_{1}\rightarrow\ell_{\alpha}\ell_{\beta},\,\nu_{\alpha}\bar{\nu}_{\beta}$
smaller with respect to the non SI extended models. In other words,
larger (smaller) contribution of the Higgs/dilaton mediated DM annihilation
channels leads to smaller (larger) contribution of the channels $N_{1}N_{1}\rightarrow\ell_{\alpha}\ell_{\beta},\,\nu_{\alpha}\bar{\nu}_{\beta}$,
and therefore smaller (larger) values of the new Yukawa couplings
$g_{i\alpha}$, i.e., more precisely the combination $\sum_{\alpha,\beta}|g_{1\alpha}g_{1\beta}^{*}|^{2}$.
Then, it would useful to define the ratio 
\begin{equation}
\mathcal{R}_{f}=\frac{\left\langle \sigma(N_{1}N_{1}\rightarrow f)\upsilon_{r}\right\rangle }{\left\langle \sigma(N_{1}N_{1})\upsilon_{r}\right\rangle },\label{eq:RL}
\end{equation}
that represents the contribution of the channel $N_{1}N_{1}\rightarrow f$
to the total thermally averaged cross section at the freeze-out. Clearly,
we have $\sum_{X}\mathcal{R}_{f}=1$, however, we will be interested
in the two ratios: $\mathcal{R}_{LL}=\sum_{\alpha.\beta}(\mathcal{R}_{\ell_{\alpha}\ell_{\beta}}+\mathcal{R}_{\nu_{\alpha}\bar{\nu}_{\beta}})$
and $\mathcal{R}_{hh}=\sum_{h_{i,k}=H,D}\mathcal{R}_{h_{i}h_{k}}$ that
represent the relative contributions of the channels $N_{1}N_{1}\rightarrow\ell_{\alpha}\ell_{\beta},\,\nu_{\alpha}\bar{\nu}_{\beta}$
and $N_{1}N_{1}\rightarrow h_{i}h_{k}$ to the annihilation cross
section, respectively. Clearly the ratio $\mathcal{R}_{LL}$ is proportional
to the combination $\sum_{\alpha,\beta}|g_{1\alpha}g_{1\beta}^{*}|^{2}$
in the limit of degenerate charged lepton masses. For a given DM mass
value $M_{1}$, let's call $\left\langle \sigma_{0}\upsilon_{r}\right\rangle =\left\langle \sigma(N_{1}N_{1})\upsilon_{r}\right\rangle $
the correct cross section value at the freeze-out that matches the
observed relic density (\ref{eq:omegah}). Then, the cross section
of all Higgs/dilaton mediated channels $\sum_{f}\left\langle \sigma(N_{1}N_{1}\rightarrow f)\upsilon_{r}\right\rangle $
($f=h_{i}h_{k},\,VV,q\bar{q}$) at the freeze-out should be smaller than
$\left\langle \sigma_{0}\upsilon_{r}\right\rangle $. Therefore, the
condition $\sum_{f}\left\langle \sigma(N_{1}N_{1}\rightarrow f)\upsilon_{r}\right\rangle \geq\left\langle \sigma_{0}\upsilon_{r}\right\rangle $
leads to a value for the relic density that is smaller than the measured
value and can not be compensated by the channels $N_{1}N_{1}\rightarrow\ell_{\alpha}\ell_{\beta},\,\nu_{\alpha}\bar{\nu}_{\beta}$.
This condition, together with the spin-independent DM DD cross section
(\ref{eq:DD}) would eliminate a significant part of the parameters
space of any SI model that addresses the neutrino mass and Majorana
DM together.

\begin{figure}[t]
\begin{centering}
\includegraphics[width=0.49\textwidth]{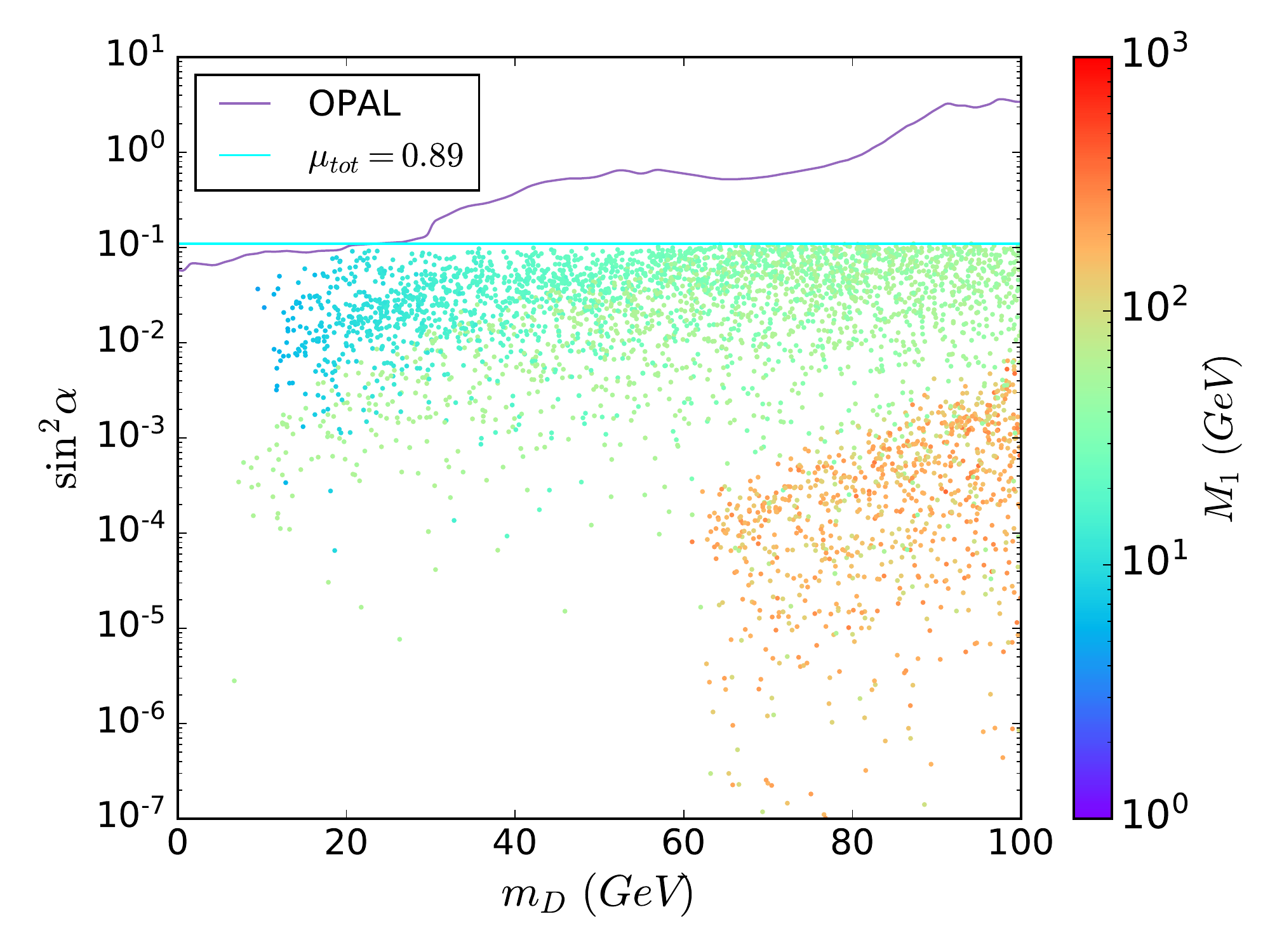}
\includegraphics[width=0.49\textwidth]{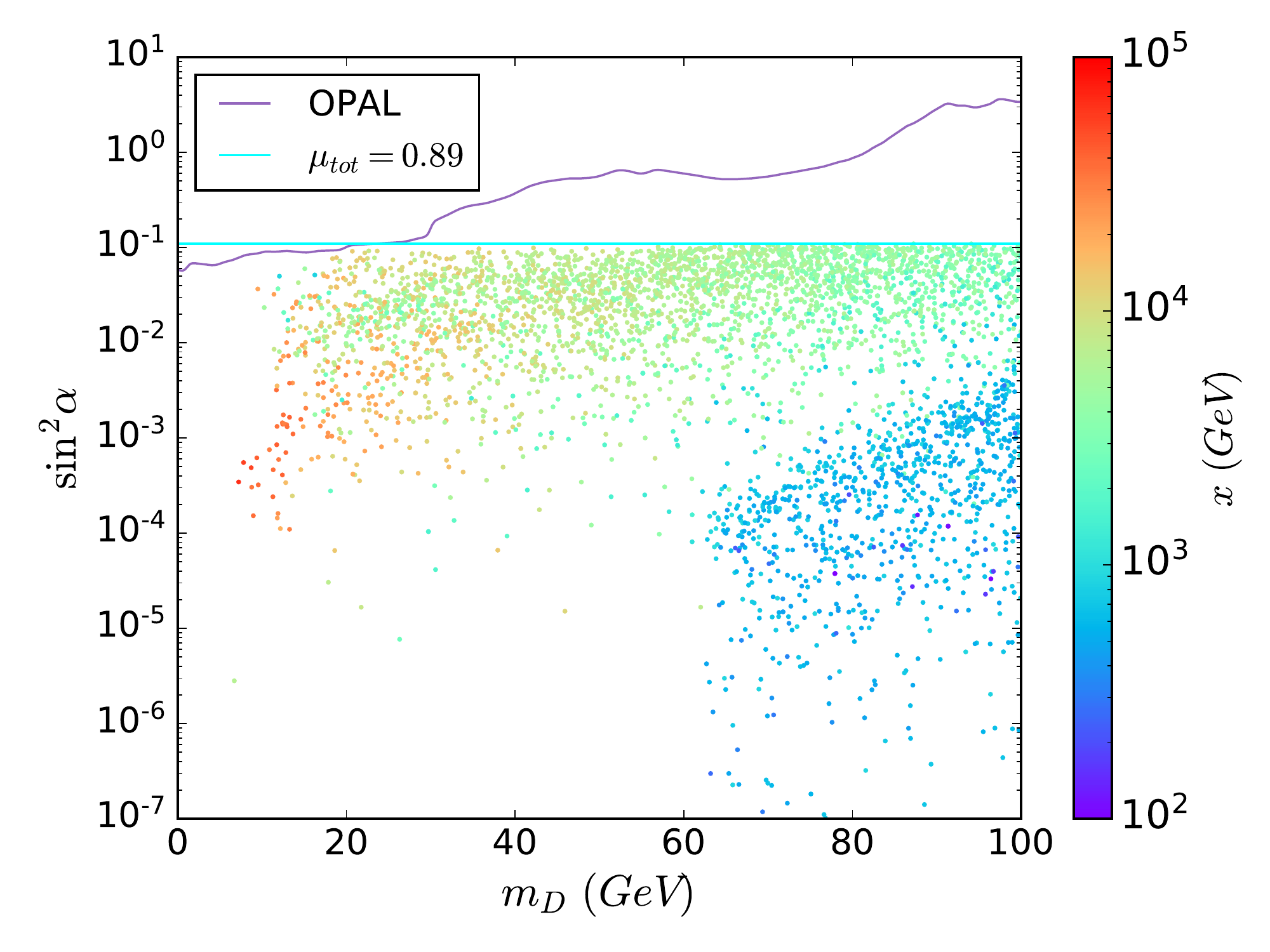}\\
 \includegraphics[width=0.49\textwidth]{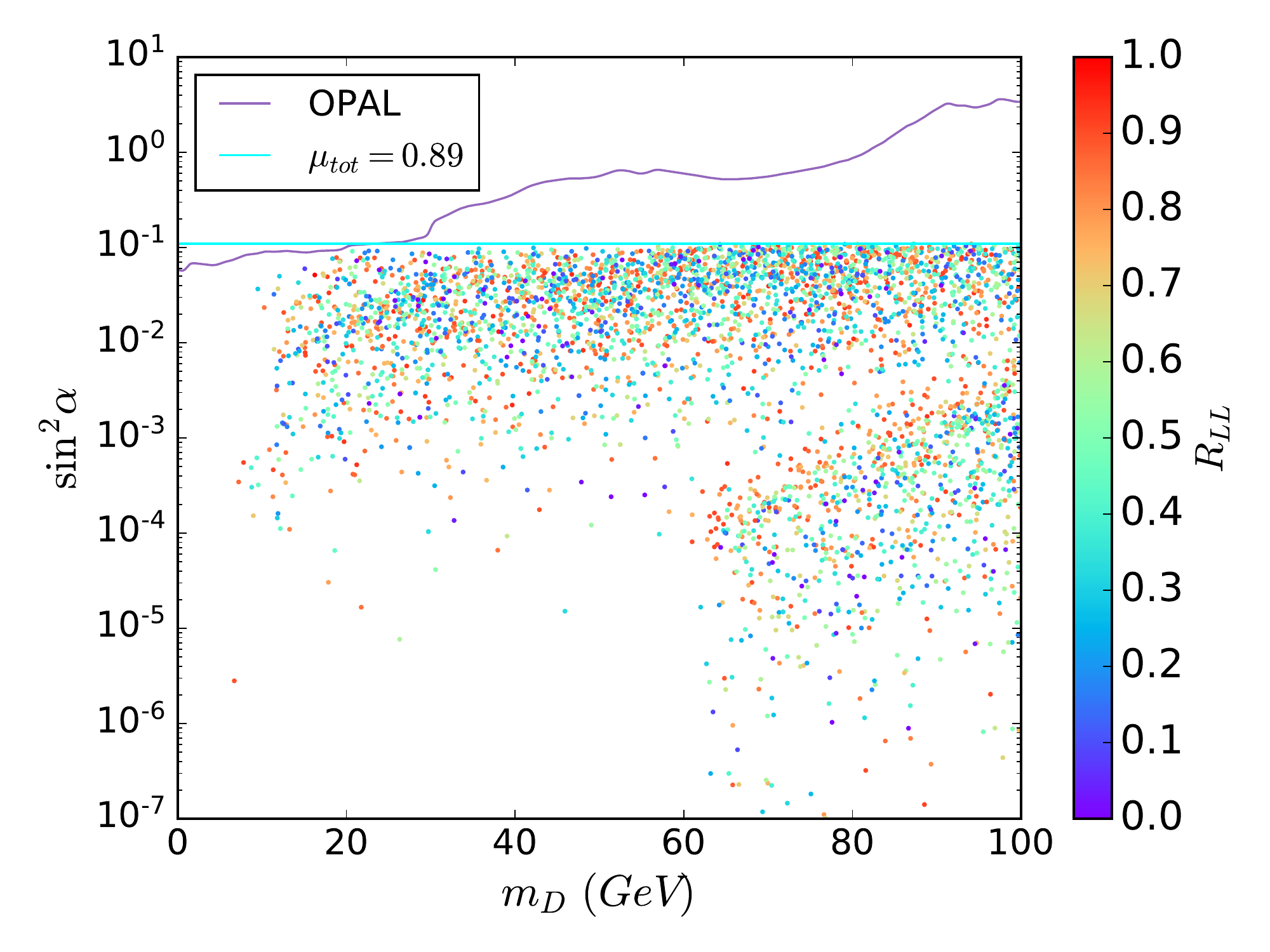}\includegraphics[width=0.49\textwidth]{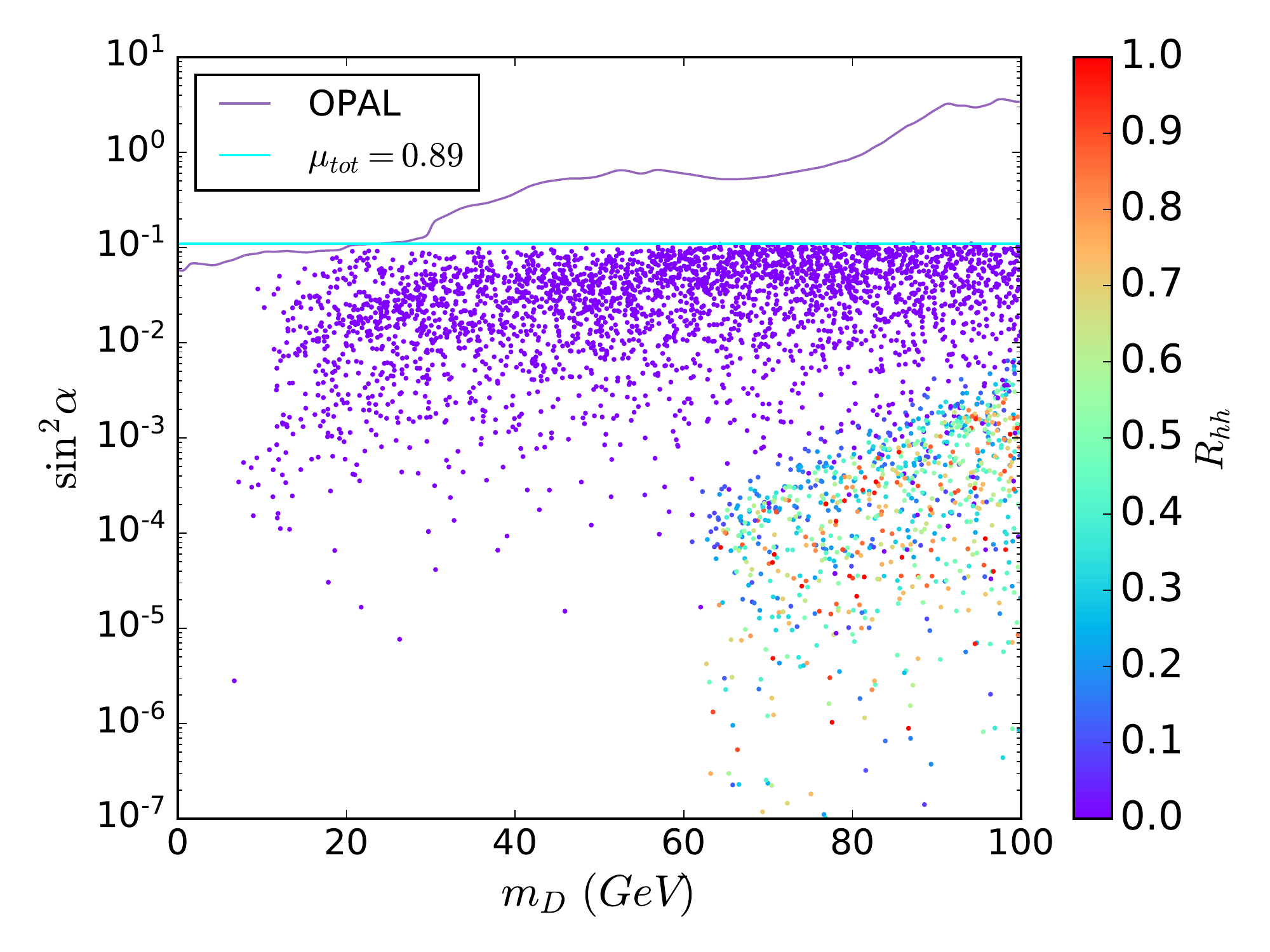}\\
\par\end{centering}
\caption{The dilaton mass versus the Higgs/dilaton mixing, where the palette
shows the DM mass (up-right), the singlet VEV (up-left), the ratio
$\mathcal{R}_{LL}$ (bottom-right) and the ratio $\mathcal{R}_{hh}$
(bottom-left). The DM requirements of relic density and direct
detection bounds, in addition to the LEP (OPAL) bound on $e^{-}e^{+}\rightarrow Z^{*}D$
are already considered.}
\label{SP} 
\end{figure}

In order to investigate the impact of these constraints, we perform
a random scan over the input parameters ranges in (\ref{eq:free}),
while taking into account the theoretical and experimental
constraints mentioned earlier. Since the case where the DM annihilation does fully occur
via the channel $N_{1}N_{1}\rightarrow\ell_{\alpha}\ell_{\beta},\,\nu_{\alpha}\bar{\nu}_{\beta}$,
is phenomenologically identical the minimal scotogenic model~\cite{Ahriche:2017iar},
we will focus on the regions of the parameters space where the contributions
of the other channels $N_{1}N_{1}\rightarrow h_{i}h_{k},\,VV,q\bar{q}$
are not vanishing. We consider 4000 benchmark points (BPs) that fulfill
all the conditions mentioned above and shown in Fig.~\ref{SP}-top
and Fig.~\ref{SP}-bottom, the possible values of \{$m_{D}-s_{\alpha}^{2}$\}
for different values of the DM mass $M_{1}$, singlet VEV $x$, and
the ratios $\mathcal{R}_{LL}$ and $\mathcal{R}_{hh}$. 

One has to mention that for a significant part the BPs (not considered
in Fig.~\ref{SP}) that pass all the requirements, the DM annihilation
is almost fully achieved via $N_{1}N_{1}\rightarrow\ell_{\alpha}\ell_{\beta},\,\nu_{\alpha}\bar{\nu}_{\beta}$.
From Fig.~\ref{SP}, one can learn many conclusions, for instance,
light dilaton ($m_{D}<25~\mathrm{GeV}$) and large mixing ($s_{\alpha}>0.1$)
correspond to light DM and large singlet VEV. In this region the DM
annihilation is dominated by light quarks, charged leptons and/or
neutrinos. Most of the BPs with large dilaton mass ($m_{D}>m_{H}/2$)
and small mixing ($s_{\alpha}<0.1$) correspond to light singlet VEV
($x<900~\mathrm{GeV}$) and heavy DM ($M_{1}\sim O(100~\mathrm{GeV})$). In this region,
the DM annihilation could be dominated either by $N_{1}N_{1}\rightarrow\ell_{\alpha}\ell_{\beta},\,\nu_{\alpha}\bar{\nu}_{\beta}$
or by the channel $N_{1}N_{1}\rightarrow S_{i}S_{k}$. For the region
of large mixing ($s_{\alpha}>0.3$) the DM annihilation channel $N_{1}N_{1}\rightarrow S_{i}S_{k}$
is always suppressed despite the dilaton mass, while the channel $N_{1}N_{1}\rightarrow\ell_{\alpha}\ell_{\beta},\,\nu_{\alpha}\bar{\nu}_{\beta}$
is dominated for part of the BPs. It should be stated that for the
region of small mixing ($s_{\alpha}<0.01$) and small dilaton mass
($m_{D}<m_{H}/2$), there exist possible viable BPs but with less
population. For these BPs, the DM annihilation is fully achieved via
the channel dominated either by $N_{1}N_{1}\rightarrow\ell_{\alpha}\ell_{\beta},\,\nu_{\alpha}\bar{\nu}_{\beta}$,
that matches phenomenologically the non-SI (minimal) scotogenic model.

\begin{figure}[t]
\begin{centering}
\includegraphics[width=0.49\textwidth]{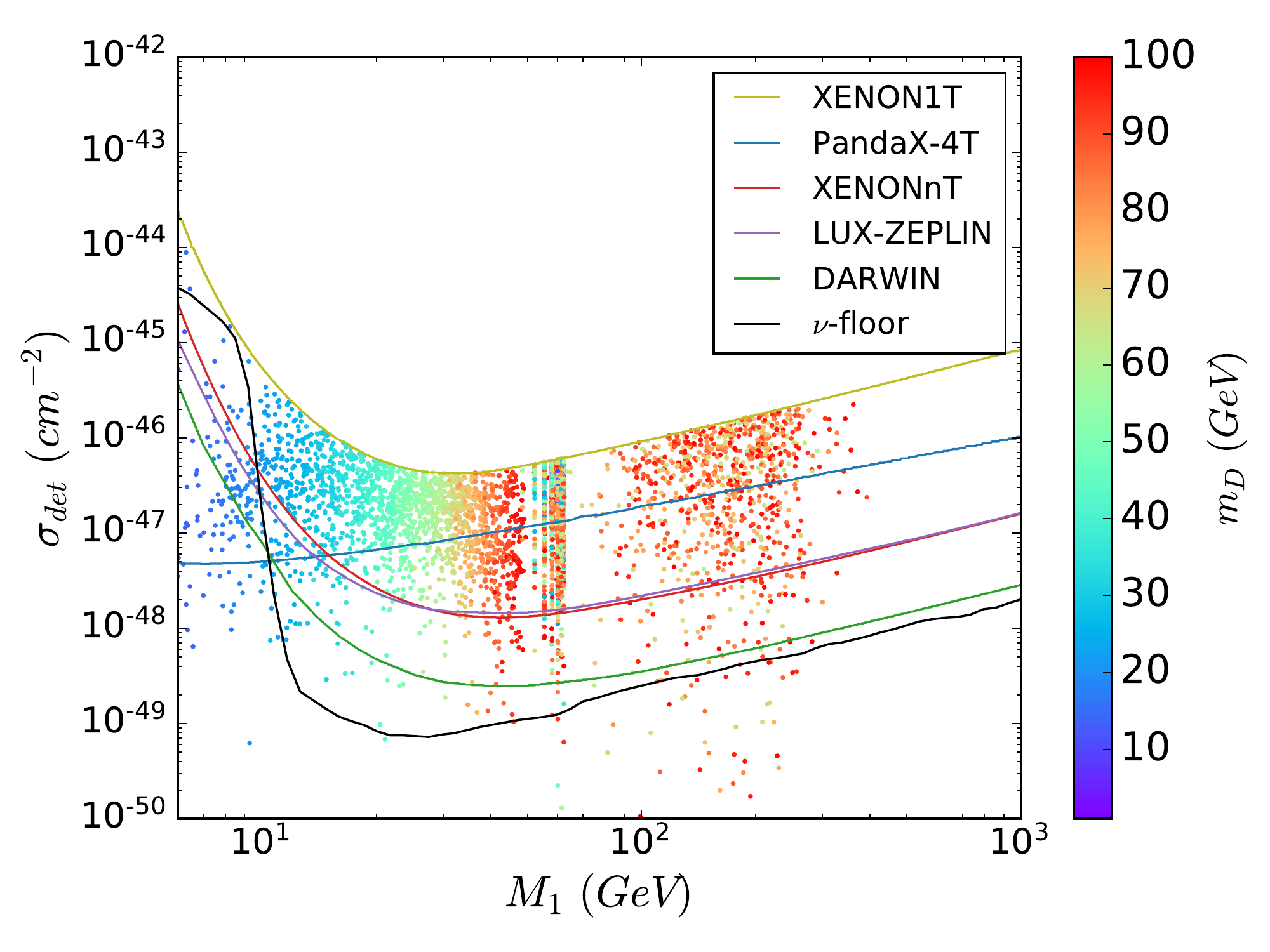}~\includegraphics[width=0.49\textwidth]{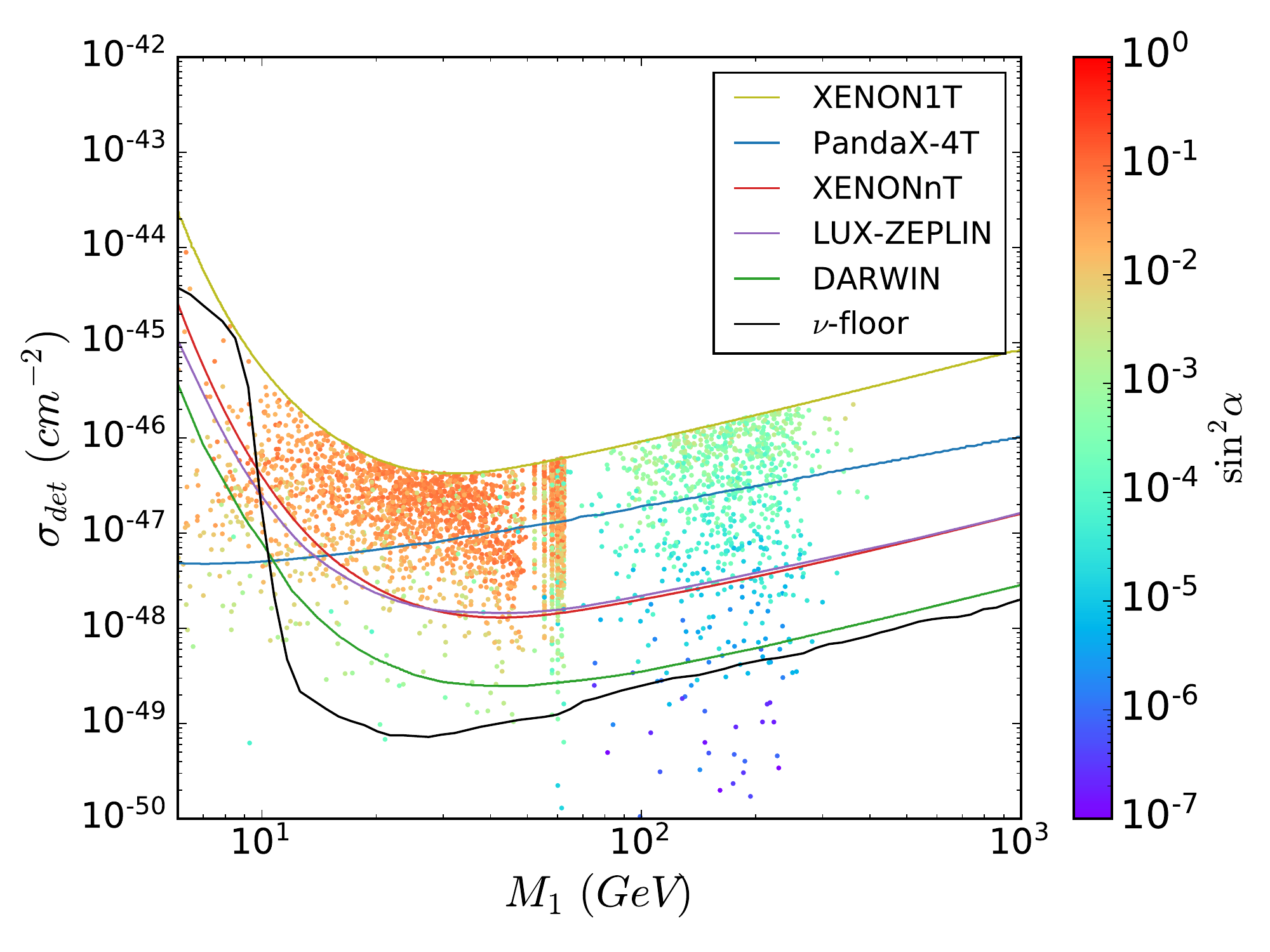} 
\par\end{centering}
\caption{The DM direct detection cross section versus the DM mass is presented, where the recent bound of Xenon 1T experiment is shown~\cite{XENON:2018voc}. the palette illustrates the dilaton mass (left) and the scalar mixing (right). 
In addition, we show the projected sensitivities for the future experiments:
PandaX-4t~\cite{PandaX:2018wtu}, LUX-Zeplin~\cite{LUX-ZEPLIN:2018poe},
XENONnT with 20 ton-yr exposure~\cite{XENON:2020kmp} and DARWIN~\cite{DARWIN:2016hyl}.
The black line represents the neutrino floor, where any of the DM detection
via nucleon scattering is not possible.}
\label{DM} 
\end{figure}

For a complete picture, we show in Fig\@.~\ref{DM} the DM-nucleon spin-independent
cross section versus the DM mass, while the palette represents the
dilaton mass (left) and the scalar mixing (right).

Clearly, from Fig.~\ref{DM} a significant part of the BPs
are within the reach of future DD experiments such as PandaX-4t~\cite{PandaX:2018wtu},
LUX-Zeplin~\cite{LUX-ZEPLIN:2018poe} and XENONnT~\cite{XENON:2020kmp},
while few of them (either with suppressed mixing $|s_{\alpha}|<10^{-3}$
or with light DM $M_{1}<10~\mathrm{GeV}$) could not before detected since
they are below the neutrino floor. For viable DM with mass values
larger than $M_{1}>100~\mathrm{GeV}$, the scalar mixing should be $|s_{\alpha}|\lesssim 0.03$
and the dilaton mass $m_{D}\gtrsim70\,\mathrm{GeV}$. However, for $20\,\mathrm{GeV}\leq M_{1}\leq60\,\mathrm{GeV}$,
the scalar mixing is almost maximal $|s_{\alpha}|\sim0.1$. The population
distribution around $M_{1}\sim m_{H}/2$ implies that the analysis
in this region requires a careful treatment.

\begin{figure}[t]
\begin{centering}
\includegraphics[width=0.49\textwidth]{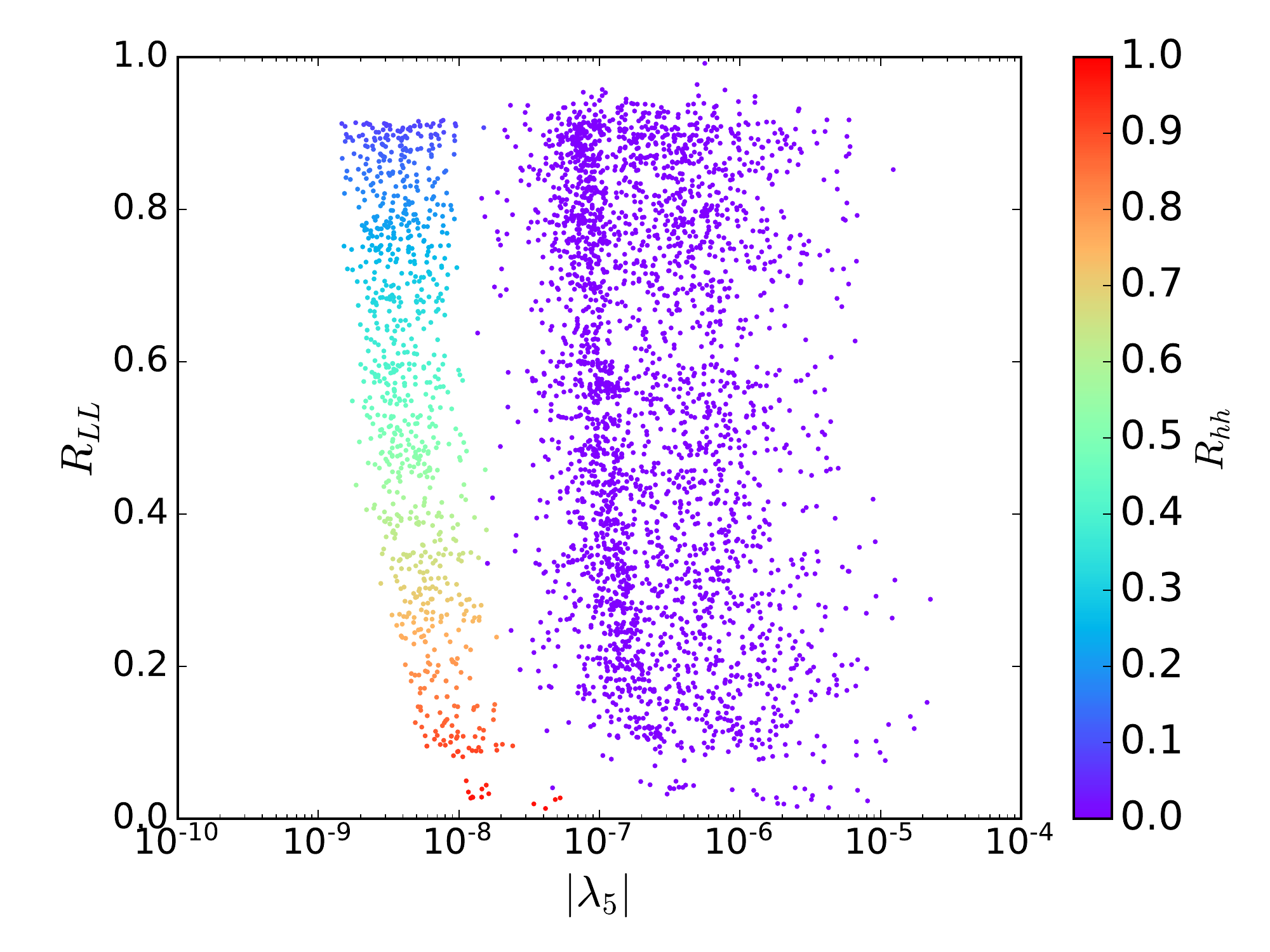}~\includegraphics[width=0.49\textwidth]{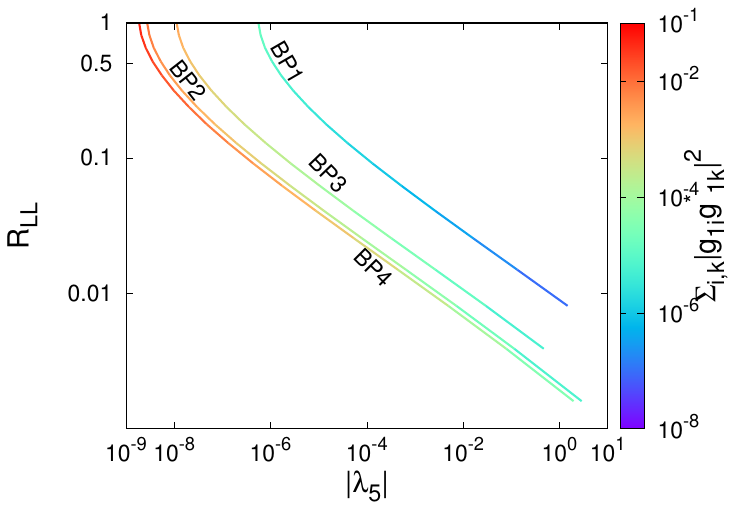} 
\par\end{centering}
\caption{Left: the ratio $\mathcal{R}_{LL}$ versus the scalar coupling $\lambda_{5}$
for the 4000 BPs used previously in Fig.~\ref{SP}, where the palette
shows the ratio $\mathcal{R}_{hh}$. Right: the dependence of the
ratio $\mathcal{R}_{LL}$ on the scalar coupling $\lambda_{5}$ for
the BPs shown in Table.~\ref{Tab}, where both DM relic density and
neutrino masses are kept in agreement with the data. The palette shows
the couplings combination $\sum_{i,k}|g_{1,i}g_{1k}^{*}|^{2}$ that
is proportional to the cross section of $N_{1}N_{1}\rightarrow\ell_{\alpha}\ell_{\beta},\,\nu_{\alpha}\bar{\nu}_{\beta}$
in the limit of massless charged leptons.}
\label{lm5} 
\end{figure}

The contribution of the channel $N_{1}N_{1}\rightarrow\ell_{\alpha}\ell_{\beta},\,\nu_{\alpha}\bar{\nu}_{\beta}$
to the total DM annihilation may take any value from almost 0\% to
almost 100\%, for the whole range of both dilation and DM masses.
This implies a relative freedom in the new Yukawa couplings range
($g_{i\alpha}$) contrary to the case of non SI version where the
new Yukawa coupling values are strictly dictated by the relic density.
In order to illustrate this point, we show in Fig.~\ref{lm5}-left
the correlation between the ratios $\mathcal{R}_{LL}$ and $\mathcal{R}_{hh}$
and the quartic coupling $\lambda_{5}$ for the 4000 BPs used previously
in Fig.~\ref{SP}. In order to show the explicit dependence of the
scalar coupling $\lambda_{5}$ on the ratio$\mathcal{R}_{LL}$, we
consider the BPs shown in Table.~\ref{Tab} among the 4000 points,
which all correspond to a maximal $\mathcal{R}_{LL}\sim1$.

\begin{table}[!h]
\begin{centering}
\begin{tabular}{|c|c|c|c|c|}
\hline 
 & BP1 & BP2 & BP3 & BP4\tabularnewline
\hline 
\hline 
$m_{D}$ (GeV) & 16.25 & 89.47 & 60.27 & 93.42\tabularnewline
\hline 
$s_{\alpha}$ & 0.22685 & 0.01467 & -0.01905 & -0.07311\tabularnewline
\hline 
x (GeV) & 856.34 & 393.7 & 1115.3 & 1657.4\tabularnewline
\hline 
$m_{S^{\pm}}$ (GeV) & 257.6 & 308.12 & 445.60 & 667.25\tabularnewline
\hline 
$\overline{m}$ (GeV) & 250.6 & 409.37 & 428.21 & 750.46\tabularnewline
\hline 
$M_{1}$ (GeV) & 6.31 & 81.94 & 175.20 & 467.81\tabularnewline
\hline 
$M_{2}$ (GeV) & 6.67 & 88.69 & 186.59 & 532.69\tabularnewline
\hline 
$M_{3}$ (GeV) & 7.34 & 96.14 & 220.03 & 602.23\tabularnewline
\hline 
$x_{f}$ & 21.94 & 24.27 & 24.89 & 25.89\tabularnewline
\hline 
$\sum_{i,k}|g_{1,i}g_{1k}^{*}|^{2}$ & $2.48\times10^{-5}$ & $3.51\times10^{-3}$ & $1.31\times10^{-2}$ & $7.02\times10^{-2}$\tabularnewline
\hline 
$\mathcal{R}_{LL}$ & 0.9918 & 0.98063 & 0.96442 & 0.95498\tabularnewline
\hline 
$\mathcal{R}_{hh}$ & $1.21\times10^{-33}$ & 0.01857 & 0.0355 & 0.0450\tabularnewline
\hline 
$\lambda_{5}$ & $5.6196\times10^{-7}$ & $1.1265\times10^{-8}$ & $2.7543\times10^{-9}$ & $-1.8715\times10^{-9}$\tabularnewline
\hline 
\end{tabular}
\par\end{centering}
\caption{The benchmark points used in Fig.~\ref{lm5}-bottom. Here, we have
$\overline{m}^{2}=\frac{1}{2}(m_{S^{0}}^{2}+m_{A^{0}}^{2})$, $x_{f}=M_{1}/T_{f}$
is freeze-out parameter, and the couplings combination $\sum_{i,k}|g_{1,i}g_{1k}^{*}|^{2}$
is proportional to the cross section of $N_{1}N_{1}\rightarrow\ell_{\alpha}\ell_{\beta},\,\nu_{\alpha}\bar{\nu}_{\beta}$ in the limit of massless charged leptons.}
\label{Tab} 
\end{table}

Next, we assume that one can make the ratio $\mathcal{R}_{LL}$ smaller
by re-scaling the new Yukawa couplings $g_{i\alpha}$ to smaller values,
while keeping the inert parameters ($M_{i},\,x,~\omega_{2},~\lambda_{3}$
and $\lambda_{4}$) constant. In this case, both neutrino mass matrix
elements in (\ref{eq:MaNU}) and the DM relic density can be kept
in agreement with the data, only by allowing $\lambda_{5}$ to have
larger values, by and tuning the radiative effects (dilaton mass and
scalar mixing), respectively. Therefore, the dependence of the ratio
$\mathcal{R}_{LL}$ on $\lambda_{5}$ is shown in Fig.~\ref{lm5}-right
where the palette shows the couplings combination $\sum_{i,k}|g_{1,i}g_{1k}^{*}|^{2}$
that is proportional to the cross section of $N_{1}N_{1}\rightarrow\ell_{\alpha}\ell_{\beta},\,\nu_{\alpha}\bar{\nu}_{\beta}$
in the massless charged leptons limit.

For the 4000 BPs considered in Fig.~\ref{lm5}-left, the
scalar coupling can be relaxed up to $\lambda_{5}\sim10^{-5}$ contrary
to the non-SI case where $\lambda_{5}\sim10^{-10}.$ From Fig.~\ref{lm5}-right,
one notices that by tuning the radiative effects (dilaton mass and
scalar mixing), one can push the scalar coupling up to larger values
as $\lambda_{5}\sim1$. Consequently, the new Yukawa couplings can
be much small as shown by the palette in Fig.~\ref{lm5}-right, i.e.,
$\lambda_{5}\sim1\Longrightarrow\sum_{i,k}|g_{1,i}g_{1k}^{*}|^{2}\leq10^{-4}$.
This conclusion can be extrapolated into other models like the SI-KNT
model~\cite{Ahriche:2015loa}, however, getting larger values for
$\lambda_{5}\sim0.1-1$ in any SI model is not possible except for
the case where the model new couplings and masses can be tuned without
being in conflict with the above mentioned theoretical and experimental
constraints. \\ 

The SI-scotogenic model with Majorana DM candidate has almost the
same LHC predictions as the minimal scotogenic model~\cite{Ahriche:2017iar}.
For instance, a pair of charged scalars that are produced at the LHC ($pp\rightarrow S^{\pm}S^{\mp}$)
leads to many distinguishable signatures such as $4\textrm{jets}+\slashed{E}_{T},~1\ell+2\textrm{jets}+\slashed{E}_{T}$
and $2\ell+\slashed{E}_{T}$. Moreover, the process $pp\rightarrow S^{\pm}S^{0},\,S^{\pm}A^{0}$ may lead to the 
signatures $1\ell+2\textrm{ jets}+\slashed{E}_{T}$ and $1\ell+\slashed{E}_{T}$. However, the process $pp\rightarrow S^{0}S^{0},\,A^{0}A^{0}$
results the final states $\slashed{E}_{T}+\textrm{ ISR}$ (mono-jet, mono-$\gamma$).
In order to distinguish the SI-scotogeneic model among its minimal version at colliders, one has to look for signatures that exit only in the
SI-version like $pp\to N_{1}N_{1}\gamma(H)$, $pp\to H\to DD\to4b,4\tau,2b2\tau,2b2\gamma$. 
As mentioned previously, the new Yukawa couplings $g_{i\alpha}$
can take small values, especially in the case where ${\cal R}_{LL} << 1$, which basically makes the life-time of the charged
scalar ($S^{\pm}$) longer and then, displaced vertices can be seen in $pp\rightarrow S^{\pm}S^{\pm},\,S^{\pm}S^{0},\,S^{\pm}A^{0}$ due to $S^{\pm}\to\ell^{\pm}N_{1}$.

\section{Conclusion\label{sec:Conc}}

In this work, we have considered the SI-scotogenic model that addresses
neutrino oscillation data, DM nature and the EWSB, all together at
the weak scale. In this model, the EWSB is triggered by radiative
effects due to new fields and interactions in addition to the SM one.
The DM candidate could be the lightest among the CP-even, CP-odd scalars
or lightest Majorana singlet fermion, here we adopted a Majorana DM
candidate. After discussing the EWSB we imposed different theoretical
and experimental constraints like the vacuum stability, the electroweak
precision tests, LEP negative searches for light scalars, the Higgs
decays ($h\rightarrow\gamma\gamma$, invisible, undertermined), DM
relic density and DM DD experiments. The DM relic density was estimated
following~\cite{Srednicki:1988ce}, where the cross section of different
annihilation channels was estimated exactly, including those of $N_{1}N_{1}\rightarrow\ell_{\alpha}\ell_{\beta},\,\nu_{\alpha}\bar{\nu}_{\beta}$.

In the non SI (minimal scotogenic) version of the model, the DM annihilation
occurs via the channel channel $N_{1}N_{1}\rightarrow\ell_{\alpha}\ell_{\beta},\,\nu_{\alpha}\bar{\nu}_{\beta}$,
which dictates the values of the new Yukawa couplings $g_{i\alpha}$
that relatively large. In addition, the neutrino mass smallness can
be achieved by imposing the a mass degeneracy between the CP-even
and the CP-odd scalars, which means a suppressed value for the coupling
$\lambda_{5}=(m_{S^{0}}^{2}-m_{A^{0}}^{2})/\upsilon^{2}\sim10^{-10}.$
Due to non negligible coupling for DM with the Higgs/dilaton, other
DM annihilation channels such as $N_{1}N_{1}\rightarrow h_{i}h_{k},\,VV,q\bar{q}$
are possible, and therefore the new Yukawa couplings $g_{i\alpha}$
could take smaller values with respect to the non-SI case. This means
that Majorana DM is possible without a mass degeneracy between the
CP-even and the CP-odd scalars.

By considering all the previous listed theoretical constraints, and
focusing on benchmark points (BPs) that are different from non SI
version, i.e., BPs with non negligible contributions of the channels
$N_{1}N_{1}\rightarrow h_{i}h_{k},\,VV,q\bar{q}$ to the total DM
annihilation, we performed a numerical scan over the parameters space.
We have found that the DM annihilation could be dominated by light
quarks, charged leptons and/or neutrinos channels for light dilaton
($m_{D}<25~\mathrm{GeV}$), large mixing ($s_{\alpha}>0.1$), light DM ($M_{1}<10\,\mathrm{GeV}$)
and large singlet VEV ($x>2\,\mathrm{TeV}$). Another interesting region for
large dilaton mass ($m_{D}>m_{H}/2$), small mixing ($s_{\alpha}<0.1$),
small singlet VEV ($x<900~\mathrm{GeV}$) and heavy DM ($M_{1}\sim O(100~\mathrm{GeV})$),
where the DM annihilation could be dominated either by $N_{1}N_{1}\rightarrow\ell_{\alpha}\ell_{\beta},\,\nu_{\alpha}\bar{\nu}_{\beta}$
or by the channel $N_{1}N_{1}\rightarrow HH,HD,DD$. We found also
that the values of the scalar quartic coupling can be relaxed up to
$\lambda_{5}\sim10^{-5}$ contrary to the non-SI case where $\lambda_{5}\sim10^{-10}.$
By possible tuning of the radiative effects (in our model due to $M_{i},\,x,~\omega_{2},~\lambda_{3}$
and $\lambda_{4}$), the scalar coupling could be $\lambda_{5}\sim1$,
which implies very small values for the new Yukawa couplings $\sum_{i,k}|g_{1,i}g_{1k}^{*}|^{2}\leq10^{-4}$.
These regions are the most important regions where the SI extension
of the scotogenic model (as well other models like the SI-KNT), in
which the Majorana DM phenomenology is different than the non SI version.
These regions are with a great interest and their phenomenology studies at
colliders is ongoing in a future work~\cite{future}.

\subsection*{Acknowledgements}

This work is supported by the University of Sharjah via the following grants: \textit{Probing the Majorana Neutrino Nature in Radiative Neutrino Mass Models at Current and Future Colliders} (No. 1802143057-P), \textit{Extended Higgs Sectors at Colliders: Constraints \& Predictions} (No. 21021430100) and \textit{Hunting for New Physics
at Colliders} (No. 21021430107). The authors thank S. Nasri for his valuable comments. 

%%%%%%%%%%%%%%%%%%%%%%%%%%%%%%%%%%%%%%

\end{document}